\documentclass[aip,apl,reprint,groupedaddress]{revtex4-1}

\usepackage{amsmath}
\usepackage{graphicx}
\graphicspath{{figures/}}
\usepackage[separate-uncertainty=true,multi-part-units=single]{siunitx}

\renewcommand{\vec}[1]{\mathbf{#1}}
\newcommand{\abs}[1]{\left\vert #1 \right\vert}

\begin{document}

\title{Holographic characterization of imperfect colloidal spheres}

\author{Mark Hannel}
\author{Christine Middleton}
\author{David G. Grier}
\affiliation{Department of Physics and Center for Soft Matter Research,
New York University, New York, NY 10003}

\begin{abstract}
We demonstrate precise measurements of the size
and refractive index of individual dimpled colloidal spheres using
holographic characterization techniques developed
for ideal spheres.
\end{abstract}

\maketitle

Holographic snapshots of colloidal spheres can be interpreted
with the Lorenz-Mie theory of light scattering to measure
an individual sphere's three-dimensional position,
size and refractive index \cite{lee07a}.
When applied to dielectric spheres with near-ideal sphericity
and smoothness, this technique yields nanometer-scale
precision for the position and radius 
\cite{lee07a,cheong10a,moyses13,krishnatreya14} 
and part-per-thousand precision for the refractive index 
\cite{lee07a,shpaisman12}.
Here, we demonstrate that this technique yields similarly precise 
and meaningful results for imperfect spheres, provided that their
deviation from sphericity is not too pronounced.

\begin{figure*}[t]
  \centering
  \includegraphics[width=0.9\textwidth]{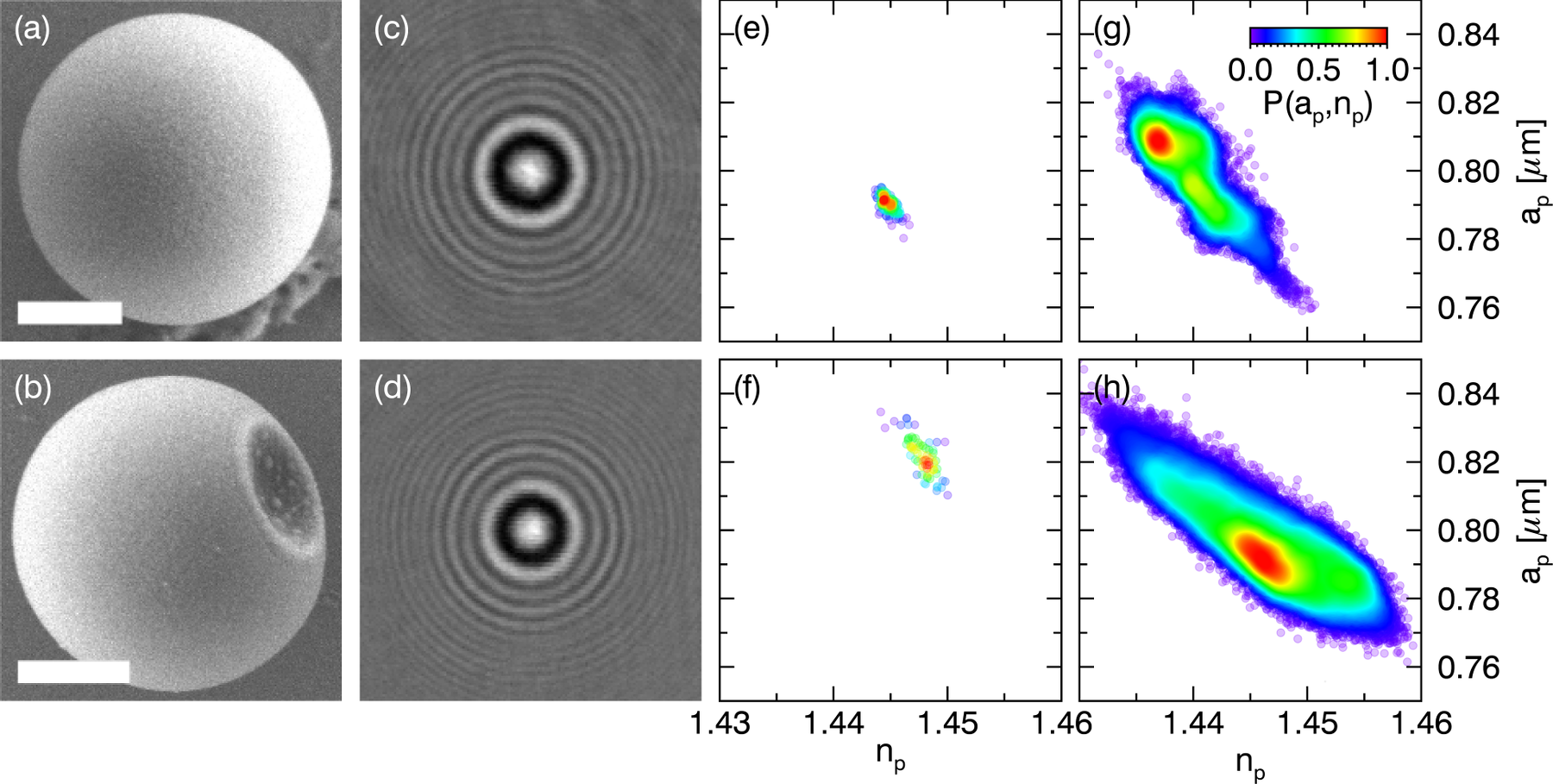}
  \caption{(a) Scanning electron micrograph of a colloidal TPM sphere and 
    (b) a dimpled sphere.  Scale bars represent \SI{500}{\nm}.
    (c) and (d) Corresponding holograms for
    particles from the samples show in (a) and (b).  (e) \num{5000} measurements
    of sphere radius and refractive index for a single
    TPM sphere held in an optical tweezer.  Each point represents
    a single measurement, and is colored according to the
    relative density of measurements, $P(a_p,n_p)$.  (f) equivalent result for
    a dimpled sphere.  (g) Distribution of size and refractive index
    for \num{5000} TPM spheres and (h) \num{5000} dimpled spheres.
   }
  \label{fig:singleparticle}
\end{figure*}

Our model system for this work consists of monodisperse
colloidal particles synthesized through emulsion polymerization of
3-methacryloxypropyl trimethoxysilane (TPM) \cite{sacanna13}.
Depending on how they are made, these particles can take the form
of spheres, as shown in Fig.~\ref{fig:singleparticle}(a), or dimpled
spheres, as shown in Fig.~\ref{fig:singleparticle}(b), in which both
the sphere radius and dimple dimensions are drawn from narrow
distributions.

The two types of particles are synthesized through similar pathways.
TPM oil, which ordinarily is insoluble in water,
undergoes hydrolysis in a basic environment (pH $> 9$) and becomes
water soluble.
Solubilized monomers then form insoluble oligomers,
which condense into spherical droplets.
Once the droplets are fully grown, they are solidified through
free radical polymerization that is initiated by adding 
$2,2^\prime$-azo-bis-isobutyrynitrile (AIBN) and 
heating to \SI{80}{\degreeCelsius} for 
\SI{2}{\hour}.
The particles then are washed and redispersed in deionized water
for study.

Homogeneously nucleated TPM droplets form spheres.
To synthesize dimpled spheres, we heterogeneously nucleate
droplet condensation by adding \SI{0.65}{\um}-diameter
polystyrene spheres to the aqueous phase.
These small spheres serve as nucleation sites for TPM condensation and remain
embedded in the surface of the resulting droplet to a depth that
depends on their wetting characteristics.
They remain in place during polymerization, and thus
determine the size of the dimple in the final particle. 
After the TPM is polymerized, the polystyrene spheres are
dissolved by transferring the particles into toluene, leaving
uniformly sized dimples.
For the sample represented by Fig.~\ref{fig:singleparticle}(b), the
dimple accounts for \SI{5}{\percent} of the equivalent sphere's volume.
The completed particles are then transferred back into 
deionized water for cleaning and study.

Dimpled spheres have immediate applications for lock-and-key
colloidal self-assembly
\cite{sacanna10,macfarlane10,sacanna11,ashton13,phillips14,wang14} 
and are models for colloidal microcapsules \cite{chang64}, which are widely
used in industrial applications, and tend to buckle into dimpled spheres
through osmotic stress \cite{chang64,knoche11,jose14}. 
They are useful for assessing the limits of holographic
characterization because their departure from sphericity is well defined.

We prepared these particles for holographic characterization
by dispersing them in water at a volume fraction of
\num{e-5} and introduced into a \SI{50}{\um}-thick gap between a
glass microscope slide and a cover slip.
The sample is sealed and mounted on the stage of a custom-built
holographic microscope \cite{lee07,lee07a}, which illuminates
it with a collimated laser beam at a vacuum wavelength
of $\lambda = \SI{447}{\nm}$ (Coherent Cube).
An illuminated particle scatters some of the laser light to the focal plane
of a microscope objective lens (Nikon Plan Apo, $100\times$, numerical
aperture 1.4, oil immersion) where it interferes with the unscattered portion
of the beam.  The interference pattern is relayed by the objective lens
and a tube lens to a video camera (NEC TI-324A), which records its intensity at
\SI{29.97}{frames\per\second} 
with a calibrated magnification of \SI{135}{\nm\per pixel}.
The camera's \SI{0.1}{\ms} exposure time is short enough to avoid
blurring of the interference pattern due to particle motion \cite{cheong09,dixon11}.
The illumination is linearly polarized with its axis of polarization aligned
to within \SI{1}{\degree} with the $\hat{x}$ axis of the camera.

Each video frame is a hologram of the particles in the
\SI{86 x 65}{\um} field of view.
The images in Fig.~\ref{fig:singleparticle}(c) and \ref{fig:singleparticle}(d)
are typical holograms of a sphere and a dimpled sphere, respectively.
They each subtend \SI[product-units=single]{150 x 150}{pixel} and 
are cropped from the camera's \SI[product-units=single]{640 x 480}{pixel} field of view.
Such holograms then can be analyzed 
\cite{lee07a,cheong09,cheong10a,krishnatreya14a} with predictions of the
Lorenz-Mie theory of light scattering \cite{bohren83,mishchenko02} 
to estimate the particle's radius, $a_p$ and refractive index, $n_p$.
Specifically, we process a recorded image $I(\vec{r})$ by subtracting off the
camera's dark count, $I_d(\vec{r})$ and normalizing by a background image,
$I_0(\vec{r})$,
that is recorded with no particles in the field of view:
\begin{equation}
  \label{eq:normalized}
  b(\vec{r}) 
  =
  \frac{I(\vec{r}) - I_d(\vec{r})}{I_0(\vec{r}) - I_d(\vec{r})}.
\end{equation}
Assuming that gradients in the amplitude and phase of the illumination
are small over the scale of the particle, the normalized hologram of
a particle located at $\vec{r}_p$ relative to the center of the microscope's
focal plane may be modeled as \cite{lee07a,krishnatreya14}
\begin{equation}
  \label{eq:lorenzmie}
  b(\vec{r}) 
  =
  \abs{
    \hat{x} + e^{-i k z_p} \, \vec{f}_s(k(\vec{r} - \vec{r}_p))
  }^2,
\end{equation}
where $k = 2 \pi n_m / \lambda$ is the wave number of the light in a medium
of refractive index $n_m$, and where $\vec{f}_s(k\vec{r})$ 
describes how the particle scatters $\hat{x}$-polarized light.

If the particle may be modeled as
an ideal isotropic sphere, then $\vec{f}_s(k\vec{r})$ is
the Lorenz-Mie scattering function \cite{bohren83,mishchenko02},
which is parameterized by the particle's radius and refractive index.
Fitting Eq.~\eqref{eq:lorenzmie} to a measured hologram
therefore yields the particle's three-dimensional position $\vec{r}_p$,
its radius $a_p$ and its refractive index $n_p$.
Previous studies on model colloidal spheres have confirmed that
these fits converge reliably for micrometer-scale spheres, and yield
the radius with a precision better than 5 nanometers 
\cite{lee07a,moyses13,krishnatreya14}
and the refractive index to within 3 parts per thousand 
\cite{shpaisman12,moyses13}.
The data in Fig.~\ref{fig:singleparticle}(e) show results obtained for
a typical TPM sphere localized in an optical trap that was
projected through the microscope's objective lens using the
holographic optical trapping technique \cite{dufresne98,*grier03}.
The spread in values is comparable to the numerically
estimated uncertainty in the individual fits,
$a_p = \SI{0.790(3)}{\um}$ and $n_p = \num{1.445(1)}$,
suggesting both that
the imaging model is appropriate, and also that the signal-to-noise
ratio estimated by the median-absolute-deviation (MAD) metric
is reasonable.
This sphere's holographically measured radius is consistent with the 
mean value, \SI{0.76(6)}{\um}, obtained through SEM observations 
on the same batch of spheres.
As expected, the polymerized particles' refractive index is larger 
than that of monomeric TPM oil, \num{1.431} at the imaging wavelength.

The scattering function for an aspherical object, 
such as a dimpled sphere, depends on the object's detailed shape
and orientation \cite{mishchenko02}.
Analytical results are available for just a few special cases.
For more general cases, numerical methods are required, 
such as the discrete dipole approximation (DDA) \cite{draine94}. 
Even with highly optimized implementations \cite{yurkin11}, however,
such approaches are computationally intensive \cite{fung12,*perry12,*wang14using}.
If an object's departure from sphericity is small enough, and if the
influence of the non-ideality on the recorded hologram is sufficiently well
localized within the recorded image, Lorenz-Mie analysis
may still yield useful results for the particle's
size and refractive index without incurring this cost.

Figure~\ref{fig:singleparticle}(f) shows results obtained by fitting the
ideal-case model to holograms of an optically trapped dimpled sphere.
The radius estimated by straightforward 
Lorenz-Mie analysis of \num{60} such holograms
is \SI{0.821(6)}{\um}.
The corresponding estimate for the refractive index,
\num{1.448(1)} is remarkably similar to the value obtained for the
ideal spheres.
In both cases, the distribution of
values extracted from nonlinear least-squares 
fits to Eq.~\eqref{eq:lorenzmie} is consistent with single-fit error estimates.
This agreement suggests that the ideal model can yield quantitative
results for the characteristics of dimpled spheres, without incurring
the costs of more realistic modeling.

SEM analysis suggests that the sample-averaged radius of the
dimpled spheres is \SI{0.75(5)}{\um}, which is significantly smaller
than the result obtained holographically.
Similar discrepancies have been noted in previous studies
\cite{yamada85,*cermola87},
and reasonably may be explained by changes
induced by preparing the spheres for SEM observation.

The extent to which a dimpled sphere's hologram may be described
with an ideal sphere's scattering function depends on the dimple's
orientation.
Optically trapping a dimpled sphere constrains its rotations 
as well as its translations, fixing the dimple's axis in the
transverse plane.
When released from its trap, the dimpled sphere rotates freely in three
dimensions, with consequences for holographic characterization.
Figure~\ref{fig:singleparticle}(g) shows the distribution of characteristics
obtained over the course of \SI{10}{\minute} for a freely diffusing
sphere.
The mean radius and refractive index obtained for this particle are
\SI{0.80(1)}{\um} and \num{1.440(3)}, respectively.
Additional uncertainty in the particle's radius and refractive index
reflects uncorrected interference artifacts in the recorded hologram due
to defects in illumination \cite{krishnatreya14}.
The corresponding distribution for the freely diffusing dimpled sphere
is broadened still further, and displays a strong anticorrelation between
radius and refractive index.
The peak of this distribution remains at the trapped-particle values,
$a_p = \SI{0.80(1)}{\um}$ and $n_p = \num{1.447(5)}$,
presumably because the randomly oriented particle is more likely 
to have its dimple transverse to the optical axis than facing it.

Having the dimple pointing sideways is beneficial for holographic
characterization.  In this orientation, the dimple's contribution to the 
scattering pattern is asymmetric, and thus minimally influences the
fit to the largely symmetric model.  It plays a role comparable to
uncorrected background artifacts, reducing the precision, but not
seriously affecting the mean values.
When the dimple is directed along the axis, however, distortions to
the scattering pattern are symmetric about the axis and thus affect
the fits more strongly.

The success of an idealized model for describing light scattering by a
dimpled sphere may be explained at least heuristically by treating
the dimple as a volume of the sphere whose scattering is phase-shifted
by \SI{180}{\degree}.
The scattering amplitudes for the sphere as a whole and for the dimple
scale roughly as their respective volumes.
If the dimple's volume is only a small fraction of the
sphere's, and if, furthermore, the dimple's contribution is asymmetric,
the perturbation should be negligible.

\begin{figure}[!t]
  \centering
  \includegraphics[width=\columnwidth]{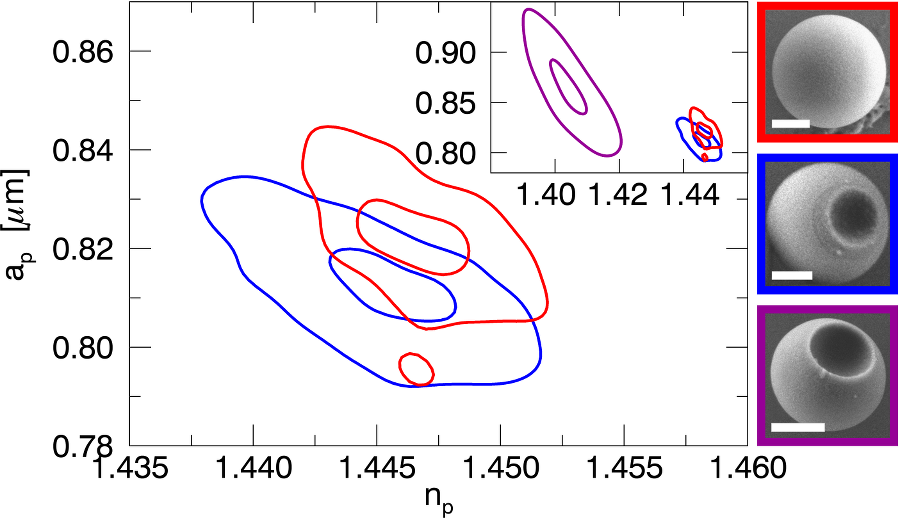}
  \caption{Level sets of the holographically measured 
    distributions of particle properties for TPM spheres
    (red), dimpled spheres with small dimples (blue) and
    large dimples (purple).  The presence of a small dimple
    has no significant influence on holographic characterization
    results.  Larger dimples cause systematic errors.
    SEM images show typical representatives of each sample,
    with scale bars denoting \SI{500}{\nm}.}
  \label{fig:distributions}
\end{figure}

We applied the same technique to characterizing the distribution
of properties in monodisperse samples of TPM spheres and 
TPM dimpled spheres.
For these measurements, dispersions of particles at a volume fraction
of \num{e-4} were streamed down a channel by a pressure driven flow
with a peak speed of $\SI{100}{\um\per\second}$.  This is fast enough
to acquire holograms of \num{5000} spheres in \SI{10}{\minute}, but
not so fast as to incur artifacts due to motion-induced blurring
\cite{cheong09,dixon11}.
Results plotted in Fig.~\ref{fig:distributions} for TPM spheres reveal
a reasonably symmetric distribution of particle sizes and refractive
indexes peaked at \SI{0.82(2)}{\um} and \num{1.446(6)}, respectively.
These values are consistent both with the single-sphere measurements
reported in Fig.~\ref{fig:singleparticle}.
The corresponding distributions for dimpled spheres plotted in
Fig.~\ref{fig:distributions} become increasingly broad and asymmetric
as the relative size of the dimple increases.
Particles with \SI{5}{\percent} dimple volumes yield a mean refractive
index, \num{1.444(6)}, consistent with the ideal spheres' value.
Increasing the dimple volume to \SI{10}{\percent} leads to
substantial deviations, with a mean refractive index of
\num{1.42(3)}.
Some variability can be attributed to the dimpled spheres'
random orientation in the channel.

\begin{figure}[!t]
  \centering
  \includegraphics[width=\columnwidth]{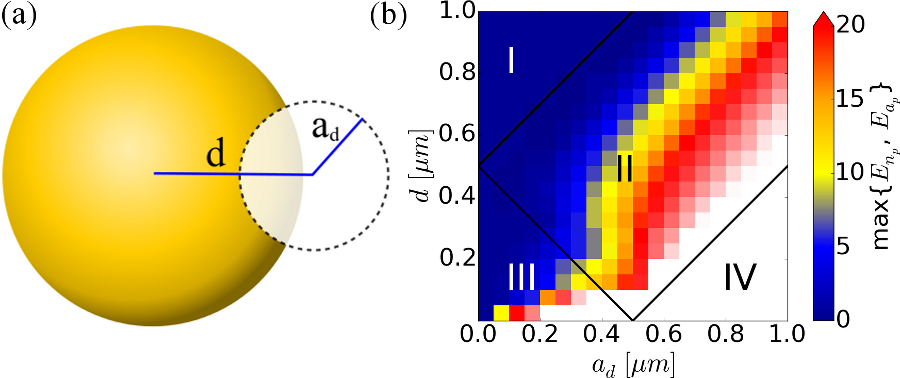}
  \caption{Holographic characterization of dimpled TPM spheres 
    with radius $a_p = \SI{0.5}{\um}$ as a function of dimple radius
    $a_d$ and inset distance, $d$.
    (a) Schematic representation of sphere geometry.
    (b) Relative error as a function of dimple radius and offset.
    Colors represent the larger of the errors 
    in particle radius and refractive index, relative to their
    respective estimated uncertainties.  Region I represents ideal
    spheres.  Region II: dimpled spheres.  Region III: spheres with
    spherical inclusions.  Region IV: inaccessible.}
  \label{fig:dimpledperformance}
\end{figure}

To assess the extent of the distortions that can be handled with the
idealized model for holographic characterization, we apply the
discrete-dipole approximation 
\cite{draine94,yurkin11,fung12,*perry12,*wang14using}
to compute holograms of 
dimpled spheres, and then analyze the resulting synthetic data
with the same software used to analyze experimental data.
The discrete-dipole approximation treats an object as a three-dimensional
arrangement of independent microscopic dipole scatterers, each of which
is illuminated by the incident beam and also by the first-order scattering
of its neighbors
\footnote{The ADDA implementation of the discrete-dipole approximation
\cite{yurkin11} used for this study discretizes the particle volume on a 
three-dimensional square grid with an effective lattice constant
roughly one-tenth the wavelength of light in the material.  Typical
numbers of dipoles range from \num{100} for the smallest particles
considered to \num{18000} for the largest.}
The superposition of scattered waves yields an estimate for 
$\vec{f}_s(k\vec{r})$, which is used to synthesize a hologram.
Dimpled spheres are modeled as the superposition of
two spheres separated by a center-to-center distance $d$, one
of radius $a_p$ and refractive index $n_p$, and the other of radius
$a_d$ and the refractive index of the medium.
This is shown schematically in Fig.~\ref{fig:dimpledperformance}(a).
Setting $a_d = 0$ or $d \ge a_p + a_d$ yields a perfect sphere.
Setting $d \le a_p - a_d$ with $a_p > a_d$ yields a sphere
with a spherical inclusion.
Dimpled spheres result when $a_p - a_d \le d \le a_p + a_d$.
Lines in Fig.~\ref{fig:dimpledperformance}(b) demarcate these
regions.

Analyzing DDA-generated holograms of perfect spheres yields 
excellent agreement
with input parameters for sphere radii up to $a_p \le \SI{0.5}{\um}$. 
Scattering by larger spheres requires proper treatment of higher-order scattering,
and is not supported by the ADDA implementation of the DDA algorithm
that we adopted \cite{yurkin11,wang14using}.
Limiting the analysis to parameters within the spherical particle's domain of
applicability, we assessed discrepancies between input parameters
and values obtained by fitting
the resulting holograms with the scattering function for ideal spheres.

Figure~\ref{fig:dimpledperformance} summarizes the performance of
Lorenz-Mie analysis for characterizing dimpled spheres.
Values in Fig.~\ref{fig:dimpledperformance}(b) represent the
factor by which the error in a dimpled particle's measured characteristics
in increased relative to the error for an ideal sphere.
As anticipated, errors in estimates for a particle's radius
and refractive index are smaller than \SI{1}{\percent} for dimples
that take up less than \SI{5}{\percent} of the sphere's volume.
Deviations become larger for spheres distorted by larger dimples,
particularly if those dimples are aligned with the optical axis.
This leaves a substantial domain of applicability within which
computationally efficient implementations of hologram analysis
can be used to measure the properties of this class of
imperfect spheres.

This work was supported primarily by the MRSEC program of the National
Science Foundation through Grant Number DMR-1420073, in part by the NSF
through Grant Number DMR-1305875, in part by
the U.S.\ Army Research Office under Grant Award No.\ W911NF-10-1-0518
and in part by a grant from Procter \& Gamble.
The holographic characterization instrument
was developed under support of the MRI program of the NSF through Grant Number
DMR-0922680.  The scanning electron microscope was purchased with financial
support from the MRI program of the NSF under Award DMR-0923251.


\begin{thebibliography}{31}%
\makeatletter
\providecommand \@ifxundefined [1]{%
 \@ifx{#1\undefined}
}%
\providecommand \@ifnum [1]{%
 \ifnum #1\expandafter \@firstoftwo
 \else \expandafter \@secondoftwo
 \fi
}%
\providecommand \@ifx [1]{%
 \ifx #1\expandafter \@firstoftwo
 \else \expandafter \@secondoftwo
 \fi
}%
\providecommand \natexlab [1]{#1}%
\providecommand \enquote  [1]{``#1''}%
\providecommand \bibnamefont  [1]{#1}%
\providecommand \bibfnamefont [1]{#1}%
\providecommand \citenamefont [1]{#1}%
\providecommand \href@noop [0]{\@secondoftwo}%
\providecommand \href [0]{\begingroup \@sanitize@url \@href}%
\providecommand \@href[1]{\@@startlink{#1}\@@href}%
\providecommand \@@href[1]{\endgroup#1\@@endlink}%
\providecommand \@sanitize@url [0]{\catcode `\\12\catcode `\$12\catcode
  `\&12\catcode `\#12\catcode `\^12\catcode `\_12\catcode `\%12\relax}%
\providecommand \@@startlink[1]{}%
\providecommand \@@endlink[0]{}%
\providecommand \url  [0]{\begingroup\@sanitize@url \@url }%
\providecommand \@url [1]{\endgroup\@href {#1}{\urlprefix }}%
\providecommand \urlprefix  [0]{URL }%
\providecommand \Eprint [0]{\href }%
\providecommand \doibase [0]{http://dx.doi.org/}%
\providecommand \selectlanguage [0]{\@gobble}%
\providecommand \bibinfo  [0]{\@secondoftwo}%
\providecommand \bibfield  [0]{\@secondoftwo}%
\providecommand \translation [1]{[#1]}%
\providecommand \BibitemOpen [0]{}%
\providecommand \bibitemStop [0]{}%
\providecommand \bibitemNoStop [0]{.\EOS\space}%
\providecommand \EOS [0]{\spacefactor3000\relax}%
\providecommand \BibitemShut  [1]{\csname bibitem#1\endcsname}%
\let\auto@bib@innerbib\@empty
\bibitem [{\citenamefont {Lee}\ \emph {et~al.}(2007)\citenamefont {Lee},
  \citenamefont {Roichman}, \citenamefont {Yi}, \citenamefont {Kim},
  \citenamefont {Yang}, \citenamefont {van Blaaderen}, \citenamefont {van
  Oostrum},\ and\ \citenamefont {Grier}}]{lee07a}%
  \BibitemOpen
  \bibfield  {author} {\bibinfo {author} {\bibfnamefont {S.-H.}\ \bibnamefont
  {Lee}}, \bibinfo {author} {\bibfnamefont {Y.}~\bibnamefont {Roichman}},
  \bibinfo {author} {\bibfnamefont {G.-R.}\ \bibnamefont {Yi}}, \bibinfo
  {author} {\bibfnamefont {S.-H.}\ \bibnamefont {Kim}}, \bibinfo {author}
  {\bibfnamefont {S.-M.}\ \bibnamefont {Yang}}, \bibinfo {author}
  {\bibfnamefont {A.}~\bibnamefont {van Blaaderen}}, \bibinfo {author}
  {\bibfnamefont {P.}~\bibnamefont {van Oostrum}}, \ and\ \bibinfo {author}
  {\bibfnamefont {D.~G.}\ \bibnamefont {Grier}},\ }\href {\doibase
  10.1364/OE.15.018275} {\bibfield  {journal} {\bibinfo  {journal} {Opt.
  Express}\ }\textbf {\bibinfo {volume} {15}},\ \bibinfo {pages} {18275}
  (\bibinfo {year} {2007})},\ \Eprint {http://arxiv.org/abs/0712.1738}
  {0712.1738} \BibitemShut {NoStop}%
\bibitem [{\citenamefont {Cheong}, \citenamefont {Krishnatreya},\ and\
  \citenamefont {Grier}(2010)}]{cheong10a}%
  \BibitemOpen
  \bibfield  {author} {\bibinfo {author} {\bibfnamefont {F.~C.}\ \bibnamefont
  {Cheong}}, \bibinfo {author} {\bibfnamefont {B.~J.}\ \bibnamefont
  {Krishnatreya}}, \ and\ \bibinfo {author} {\bibfnamefont {D.~G.}\
  \bibnamefont {Grier}},\ }\href {\doibase 10.1364/OE.18.013563} {\bibfield
  {journal} {\bibinfo  {journal} {Opt. Express}\ }\textbf {\bibinfo {volume}
  {18}},\ \bibinfo {pages} {13563} (\bibinfo {year} {2010})}\BibitemShut
  {NoStop}%
\bibitem [{\citenamefont {Moyses}, \citenamefont {Krishnatreya},\ and\
  \citenamefont {Grier}(2013)}]{moyses13}%
  \BibitemOpen
  \bibfield  {author} {\bibinfo {author} {\bibfnamefont {H.}~\bibnamefont
  {Moyses}}, \bibinfo {author} {\bibfnamefont {B.~J.}\ \bibnamefont
  {Krishnatreya}}, \ and\ \bibinfo {author} {\bibfnamefont {D.~G.}\
  \bibnamefont {Grier}},\ }\href {\doibase 10.1364/OE.21.005968} {\bibfield
  {journal} {\bibinfo  {journal} {Opt. Express}\ }\textbf {\bibinfo {volume}
  {21}},\ \bibinfo {pages} {5968} (\bibinfo {year} {2013})}\BibitemShut
  {NoStop}%
\bibitem [{\citenamefont {Krishnatreya}\ \emph {et~al.}(2014)\citenamefont
  {Krishnatreya}, \citenamefont {Colen-Landy}, \citenamefont {Hasebe},
  \citenamefont {Bell}, \citenamefont {Jones}, \citenamefont {Sunda-Meya},\
  and\ \citenamefont {Grier}}]{krishnatreya14}%
  \BibitemOpen
  \bibfield  {author} {\bibinfo {author} {\bibfnamefont {B.~J.}\ \bibnamefont
  {Krishnatreya}}, \bibinfo {author} {\bibfnamefont {A.}~\bibnamefont
  {Colen-Landy}}, \bibinfo {author} {\bibfnamefont {P.}~\bibnamefont {Hasebe}},
  \bibinfo {author} {\bibfnamefont {B.~A.}\ \bibnamefont {Bell}}, \bibinfo
  {author} {\bibfnamefont {J.~R.}\ \bibnamefont {Jones}}, \bibinfo {author}
  {\bibfnamefont {A.}~\bibnamefont {Sunda-Meya}}, \ and\ \bibinfo {author}
  {\bibfnamefont {D.~G.}\ \bibnamefont {Grier}},\ }\href {\doibase
  10.1119/1.4827275} {\bibfield  {journal} {\bibinfo  {journal} {Am. J. Phys.}\
  }\textbf {\bibinfo {volume} {82}},\ \bibinfo {pages} {23} (\bibinfo {year}
  {2014})}\BibitemShut {NoStop}%
\bibitem [{\citenamefont {Shpaisman}, \citenamefont {Krishnatreya},\ and\
  \citenamefont {Grier}(2012)}]{shpaisman12}%
  \BibitemOpen
  \bibfield  {author} {\bibinfo {author} {\bibfnamefont {H.}~\bibnamefont
  {Shpaisman}}, \bibinfo {author} {\bibfnamefont {B.~J.}\ \bibnamefont
  {Krishnatreya}}, \ and\ \bibinfo {author} {\bibfnamefont {D.~G.}\
  \bibnamefont {Grier}},\ }\href {\doibase 10.1063/1.4747168} {\bibfield
  {journal} {\bibinfo  {journal} {Appl. Phys. Lett.}\ }\textbf {\bibinfo
  {volume} {101}},\ \bibinfo {pages} {091102} (\bibinfo {year}
  {2012})}\BibitemShut {NoStop}%
\bibitem [{\citenamefont {Sacanna}\ \emph {et~al.}(2013)\citenamefont
  {Sacanna}, \citenamefont {Korpics}, \citenamefont {Rodriguez}, \citenamefont
  {Colon-Melendez}, \citenamefont {Kim}, \citenamefont {Pine},\ and\
  \citenamefont {Yi}}]{sacanna13}%
  \BibitemOpen
  \bibfield  {author} {\bibinfo {author} {\bibfnamefont {S.}~\bibnamefont
  {Sacanna}}, \bibinfo {author} {\bibfnamefont {M.}~\bibnamefont {Korpics}},
  \bibinfo {author} {\bibfnamefont {K.}~\bibnamefont {Rodriguez}}, \bibinfo
  {author} {\bibfnamefont {L.}~\bibnamefont {Colon-Melendez}}, \bibinfo
  {author} {\bibfnamefont {S.-H.}\ \bibnamefont {Kim}}, \bibinfo {author}
  {\bibfnamefont {D.~J.}\ \bibnamefont {Pine}}, \ and\ \bibinfo {author}
  {\bibfnamefont {G.-R.}\ \bibnamefont {Yi}},\ }\href@noop {} {\bibfield
  {journal} {\bibinfo  {journal} {Nature Commun.}\ }\textbf {\bibinfo {volume}
  {4}},\ \bibinfo {pages} {1688} (\bibinfo {year} {2013})}\BibitemShut
  {NoStop}%
\bibitem [{\citenamefont {Sacanna}\ \emph {et~al.}(2010)\citenamefont
  {Sacanna}, \citenamefont {Irvine}, \citenamefont {Chaikin},\ and\
  \citenamefont {Pine}}]{sacanna10}%
  \BibitemOpen
  \bibfield  {author} {\bibinfo {author} {\bibfnamefont {S.}~\bibnamefont
  {Sacanna}}, \bibinfo {author} {\bibfnamefont {W.~T.~M.}\ \bibnamefont
  {Irvine}}, \bibinfo {author} {\bibfnamefont {P.~M.}\ \bibnamefont {Chaikin}},
  \ and\ \bibinfo {author} {\bibfnamefont {D.~J.}\ \bibnamefont {Pine}},\
  }\href@noop {} {\bibfield  {journal} {\bibinfo  {journal} {Nature}\ }\textbf
  {\bibinfo {volume} {464}},\ \bibinfo {pages} {575} (\bibinfo {year}
  {2010})}\BibitemShut {NoStop}%
\bibitem [{\citenamefont {Macfarlane}\ and\ \citenamefont
  {Mirkin}(2010)}]{macfarlane10}%
  \BibitemOpen
  \bibfield  {author} {\bibinfo {author} {\bibfnamefont {J.}~\bibnamefont
  {Macfarlane}, \bibfnamefont {Robert}}\ and\ \bibinfo {author} {\bibfnamefont
  {C.~A.}\ \bibnamefont {Mirkin}},\ }\href@noop {} {\bibfield  {journal}
  {\bibinfo  {journal} {ChemPhysChem}\ }\textbf {\bibinfo {volume} {11}},\
  \bibinfo {pages} {3215} (\bibinfo {year} {2010})}\BibitemShut {NoStop}%
\bibitem [{\citenamefont {Sacanna}\ \emph {et~al.}(2011)\citenamefont
  {Sacanna}, \citenamefont {Irvine}, \citenamefont {Rossi},\ and\ \citenamefont
  {Pine}}]{sacanna11}%
  \BibitemOpen
  \bibfield  {author} {\bibinfo {author} {\bibfnamefont {S.}~\bibnamefont
  {Sacanna}}, \bibinfo {author} {\bibfnamefont {W.~T.~M.}\ \bibnamefont
  {Irvine}}, \bibinfo {author} {\bibfnamefont {L.}~\bibnamefont {Rossi}}, \
  and\ \bibinfo {author} {\bibfnamefont {D.~J.}\ \bibnamefont {Pine}},\
  }\href@noop {} {\bibfield  {journal} {\bibinfo  {journal} {Superlattices
  Microstructures}\ }\textbf {\bibinfo {volume} {7}},\ \bibinfo {pages} {1631}
  (\bibinfo {year} {2011})}\BibitemShut {NoStop}%
\bibitem [{\citenamefont {Ashton}, \citenamefont {Jack},\ and\ \citenamefont
  {Wilding}(2013)}]{ashton13}%
  \BibitemOpen
  \bibfield  {author} {\bibinfo {author} {\bibfnamefont {D.~J.}\ \bibnamefont
  {Ashton}}, \bibinfo {author} {\bibfnamefont {R.~L.}\ \bibnamefont {Jack}}, \
  and\ \bibinfo {author} {\bibfnamefont {N.~B.}\ \bibnamefont {Wilding}},\
  }\href@noop {} {\bibfield  {journal} {\bibinfo  {journal} {Superlattices
  Microstructures}\ }\textbf {\bibinfo {volume} {9}},\ \bibinfo {pages} {9661}
  (\bibinfo {year} {2013})}\BibitemShut {NoStop}%
\bibitem [{\citenamefont {Phillips}\ \emph {et~al.}(2014)\citenamefont
  {Phillips}, \citenamefont {Jankowski}, \citenamefont {Krishnatreya},
  \citenamefont {Edmond}, \citenamefont {Sacanna}, \citenamefont {Grier},
  \citenamefont {Pine},\ and\ \citenamefont {Glotzer}}]{phillips14}%
  \BibitemOpen
  \bibfield  {author} {\bibinfo {author} {\bibfnamefont {C.~L.}\ \bibnamefont
  {Phillips}}, \bibinfo {author} {\bibfnamefont {E.}~\bibnamefont {Jankowski}},
  \bibinfo {author} {\bibfnamefont {B.~J.}\ \bibnamefont {Krishnatreya}},
  \bibinfo {author} {\bibfnamefont {K.~V.}\ \bibnamefont {Edmond}}, \bibinfo
  {author} {\bibfnamefont {S.}~\bibnamefont {Sacanna}}, \bibinfo {author}
  {\bibfnamefont {D.~G.}\ \bibnamefont {Grier}}, \bibinfo {author}
  {\bibfnamefont {D.~J.}\ \bibnamefont {Pine}}, \ and\ \bibinfo {author}
  {\bibfnamefont {S.~C.}\ \bibnamefont {Glotzer}},\ }\href {\doibase
  10.1039/C4SM00796D} {\bibfield  {journal} {\bibinfo  {journal} {Soft Matter}\
  }\textbf {\bibinfo {volume} {10}},\ \bibinfo {pages} {7395} (\bibinfo {year}
  {2014})}\BibitemShut {NoStop}%
\bibitem [{\citenamefont {Wang}\ \emph
  {et~al.}(2014{\natexlab{a}})\citenamefont {Wang}, \citenamefont {Wang},
  \citenamefont {Zheng}, \citenamefont {Yi}, \citenamefont {Sacanna},
  \citenamefont {Pine},\ and\ \citenamefont {Weck}}]{wang14}%
  \BibitemOpen
  \bibfield  {author} {\bibinfo {author} {\bibfnamefont {Y.}~\bibnamefont
  {Wang}}, \bibinfo {author} {\bibfnamefont {Y.}~\bibnamefont {Wang}}, \bibinfo
  {author} {\bibfnamefont {X.}~\bibnamefont {Zheng}}, \bibinfo {author}
  {\bibfnamefont {G.-R.}\ \bibnamefont {Yi}}, \bibinfo {author} {\bibfnamefont
  {S.}~\bibnamefont {Sacanna}}, \bibinfo {author} {\bibfnamefont {D.~J.}\
  \bibnamefont {Pine}}, \ and\ \bibinfo {author} {\bibfnamefont
  {M.}~\bibnamefont {Weck}},\ }\href@noop {} {\bibfield  {journal} {\bibinfo
  {journal} {J. Am. Chem. Soc.}\ }\textbf {\bibinfo {volume} {136}},\ \bibinfo
  {pages} {6866} (\bibinfo {year} {2014}{\natexlab{a}})}\BibitemShut {NoStop}%
\bibitem [{\citenamefont {Chang}(1964)}]{chang64}%
  \BibitemOpen
  \bibfield  {author} {\bibinfo {author} {\bibfnamefont {T.~M.~S.}\
  \bibnamefont {Chang}},\ }\href@noop {} {\bibfield  {journal} {\bibinfo
  {journal} {Science}\ }\textbf {\bibinfo {volume} {146}},\ \bibinfo {pages}
  {524} (\bibinfo {year} {1964})}\BibitemShut {NoStop}%
\bibitem [{\citenamefont {Knoche}\ and\ \citenamefont
  {Kierfeld}(2011)}]{knoche11}%
  \BibitemOpen
  \bibfield  {author} {\bibinfo {author} {\bibfnamefont {S.}~\bibnamefont
  {Knoche}}\ and\ \bibinfo {author} {\bibfnamefont {J.}~\bibnamefont
  {Kierfeld}},\ }\href@noop {} {\bibfield  {journal} {\bibinfo  {journal}
  {Phys. Rev. E}\ }\textbf {\bibinfo {volume} {84}},\ \bibinfo {pages} {046608}
  (\bibinfo {year} {2011})}\BibitemShut {NoStop}%
\bibitem [{\citenamefont {Jose}\ \emph {et~al.}(2014)\citenamefont {Jose},
  \citenamefont {Kamp}, \citenamefont {van Blaaderen},\ and\ \citenamefont
  {Imhof}}]{jose14}%
  \BibitemOpen
  \bibfield  {author} {\bibinfo {author} {\bibfnamefont {J.}~\bibnamefont
  {Jose}}, \bibinfo {author} {\bibfnamefont {M.}~\bibnamefont {Kamp}}, \bibinfo
  {author} {\bibfnamefont {A.}~\bibnamefont {van Blaaderen}}, \ and\ \bibinfo
  {author} {\bibfnamefont {A.}~\bibnamefont {Imhof}},\ }\href@noop {}
  {\bibfield  {journal} {\bibinfo  {journal} {Langmuir}\ }\textbf {\bibinfo
  {volume} {30}},\ \bibinfo {pages} {2385} (\bibinfo {year}
  {2014})}\BibitemShut {NoStop}%
\bibitem [{\citenamefont {Lee}\ and\ \citenamefont {Grier}(2007)}]{lee07}%
  \BibitemOpen
  \bibfield  {author} {\bibinfo {author} {\bibfnamefont {S.-H.}\ \bibnamefont
  {Lee}}\ and\ \bibinfo {author} {\bibfnamefont {D.~G.}\ \bibnamefont
  {Grier}},\ }\href {\doibase 10.1364/OE.15.001505} {\bibfield  {journal}
  {\bibinfo  {journal} {Opt. Express}\ }\textbf {\bibinfo {volume} {15}},\
  \bibinfo {pages} {1505} (\bibinfo {year} {2007})}\BibitemShut {NoStop}%
\bibitem [{\citenamefont {Cheong}\ \emph {et~al.}(2009)\citenamefont {Cheong},
  \citenamefont {Sun}, \citenamefont {Dreyfus}, \citenamefont {Amato-Grill},
  \citenamefont {Xiao}, \citenamefont {Dixon},\ and\ \citenamefont
  {Grier}}]{cheong09}%
  \BibitemOpen
  \bibfield  {author} {\bibinfo {author} {\bibfnamefont {F.~C.}\ \bibnamefont
  {Cheong}}, \bibinfo {author} {\bibfnamefont {B.}~\bibnamefont {Sun}},
  \bibinfo {author} {\bibfnamefont {R.}~\bibnamefont {Dreyfus}}, \bibinfo
  {author} {\bibfnamefont {J.}~\bibnamefont {Amato-Grill}}, \bibinfo {author}
  {\bibfnamefont {K.}~\bibnamefont {Xiao}}, \bibinfo {author} {\bibfnamefont
  {L.}~\bibnamefont {Dixon}}, \ and\ \bibinfo {author} {\bibfnamefont {D.~G.}\
  \bibnamefont {Grier}},\ }\href {\doibase 10.1364/OE.17.013071} {\bibfield
  {journal} {\bibinfo  {journal} {Opt. Express}\ }\textbf {\bibinfo {volume}
  {17}},\ \bibinfo {pages} {13071} (\bibinfo {year} {2009})}\BibitemShut
  {NoStop}%
\bibitem [{\citenamefont {Dixon}, \citenamefont {Cheong},\ and\ \citenamefont
  {Grier}(2011)}]{dixon11}%
  \BibitemOpen
  \bibfield  {author} {\bibinfo {author} {\bibfnamefont {L.}~\bibnamefont
  {Dixon}}, \bibinfo {author} {\bibfnamefont {F.~C.}\ \bibnamefont {Cheong}}, \
  and\ \bibinfo {author} {\bibfnamefont {D.~G.}\ \bibnamefont {Grier}},\ }\href
  {\doibase 10.1364/OE.19.004393} {\bibfield  {journal} {\bibinfo  {journal}
  {Opt. Express}\ }\textbf {\bibinfo {volume} {19}},\ \bibinfo {pages} {4393}
  (\bibinfo {year} {2011})}\BibitemShut {NoStop}%
\bibitem [{\citenamefont {Krishnatreya}\ and\ \citenamefont
  {Grier}(2014)}]{krishnatreya14a}%
  \BibitemOpen
  \bibfield  {author} {\bibinfo {author} {\bibfnamefont {B.~J.}\ \bibnamefont
  {Krishnatreya}}\ and\ \bibinfo {author} {\bibfnamefont {D.~G.}\ \bibnamefont
  {Grier}},\ }\href {\doibase 10.1364/OE.22.012773} {\bibfield  {journal}
  {\bibinfo  {journal} {Opt. Express}\ }\textbf {\bibinfo {volume} {22}},\
  \bibinfo {pages} {12773} (\bibinfo {year} {2014})}\BibitemShut {NoStop}%
\bibitem [{\citenamefont {Bohren}\ and\ \citenamefont
  {Huffman}(1983)}]{bohren83}%
  \BibitemOpen
  \bibfield  {author} {\bibinfo {author} {\bibfnamefont {C.~F.}\ \bibnamefont
  {Bohren}}\ and\ \bibinfo {author} {\bibfnamefont {D.~R.}\ \bibnamefont
  {Huffman}},\ }\href@noop {} {\emph {\bibinfo {title} {Absorption and
  Scattering of Light by Small Particles}}}\ (\bibinfo  {publisher} {Wiley
  Interscience},\ \bibinfo {address} {New York},\ \bibinfo {year}
  {1983})\BibitemShut {NoStop}%
\bibitem [{\citenamefont {Mishchenko}, \citenamefont {Travis},\ and\
  \citenamefont {Lacis}(2001)}]{mishchenko02}%
  \BibitemOpen
  \bibfield  {author} {\bibinfo {author} {\bibfnamefont {M.~I.}\ \bibnamefont
  {Mishchenko}}, \bibinfo {author} {\bibfnamefont {L.~D.}\ \bibnamefont
  {Travis}}, \ and\ \bibinfo {author} {\bibfnamefont {A.~A.}\ \bibnamefont
  {Lacis}},\ }\href@noop {} {\emph {\bibinfo {title} {Scattering, Absorption
  and Emission of Light by Small Particles}}}\ (\bibinfo  {publisher}
  {Cambridge University Press},\ \bibinfo {address} {Cambridge},\ \bibinfo
  {year} {2001})\BibitemShut {NoStop}%
\bibitem [{\citenamefont {Dufresne}\ and\ \citenamefont
  {Grier}(1998)}]{dufresne98}%
  \BibitemOpen
  \bibfield  {author} {\bibinfo {author} {\bibfnamefont {E.~R.}\ \bibnamefont
  {Dufresne}}\ and\ \bibinfo {author} {\bibfnamefont {D.~G.}\ \bibnamefont
  {Grier}},\ }\href {\doibase 10.1063/1.1148883} {\bibfield  {journal}
  {\bibinfo  {journal} {Rev. Sci. Instrum.}\ }\textbf {\bibinfo {volume}
  {69}},\ \bibinfo {pages} {1974} (\bibinfo {year} {1998})}\BibitemShut
  {NoStop}%
\bibitem [{\citenamefont {Grier}(2003)}]{grier03}%
  \BibitemOpen
  \bibfield  {author} {\bibinfo {author} {\bibfnamefont {D.~G.}\ \bibnamefont
  {Grier}},\ }\href {\doibase 10.1038/nature01935} {\bibfield  {journal}
  {\bibinfo  {journal} {Nature}\ }\textbf {\bibinfo {volume} {424}},\ \bibinfo
  {pages} {810} (\bibinfo {year} {2003})}\BibitemShut {NoStop}%
\bibitem [{\citenamefont {Draine}\ and\ \citenamefont
  {Flatau}(1994)}]{draine94}%
  \BibitemOpen
  \bibfield  {author} {\bibinfo {author} {\bibfnamefont {B.~T.}\ \bibnamefont
  {Draine}}\ and\ \bibinfo {author} {\bibfnamefont {P.~J.}\ \bibnamefont
  {Flatau}},\ }\href@noop {} {\bibfield  {journal} {\bibinfo  {journal} {J.
  Opt. Soc. Am. A}\ }\textbf {\bibinfo {volume} {11}},\ \bibinfo {pages} {1491}
  (\bibinfo {year} {1994})}\BibitemShut {NoStop}%
\bibitem [{\citenamefont {Yurkin}\ and\ \citenamefont
  {Hoekstra}(2011)}]{yurkin11}%
  \BibitemOpen
  \bibfield  {author} {\bibinfo {author} {\bibfnamefont {M.~A.}\ \bibnamefont
  {Yurkin}}\ and\ \bibinfo {author} {\bibfnamefont {A.~G.}\ \bibnamefont
  {Hoekstra}},\ }\href@noop {} {\bibfield  {journal} {\bibinfo  {journal} {J.
  Quant. Spectr. Rad. Trans.}\ }\textbf {\bibinfo {volume} {112}},\ \bibinfo
  {pages} {2234} (\bibinfo {year} {2011})}\BibitemShut {NoStop}%
\bibitem [{\citenamefont {Fung}\ \emph {et~al.}(2012)\citenamefont {Fung},
  \citenamefont {Perry}, \citenamefont {Dimiduk},\ and\ \citenamefont
  {Manoharan}}]{fung12}%
  \BibitemOpen
  \bibfield  {author} {\bibinfo {author} {\bibfnamefont {J.}~\bibnamefont
  {Fung}}, \bibinfo {author} {\bibfnamefont {R.~W.}\ \bibnamefont {Perry}},
  \bibinfo {author} {\bibfnamefont {T.~G.}\ \bibnamefont {Dimiduk}}, \ and\
  \bibinfo {author} {\bibfnamefont {V.~N.}\ \bibnamefont {Manoharan}},\
  }\href@noop {} {\bibfield  {journal} {\bibinfo  {journal} {J. Quant. Spectr.
  Rad. Trans.}\ }\textbf {\bibinfo {volume} {113}},\ \bibinfo {pages} {212}
  (\bibinfo {year} {2012})}\BibitemShut {NoStop}%
\bibitem [{\citenamefont {Perry}\ \emph {et~al.}(2012)\citenamefont {Perry},
  \citenamefont {Meng}, \citenamefont {Dimiduk}, \citenamefont {Fung},\ and\
  \citenamefont {Manoharan}}]{perry12}%
  \BibitemOpen
  \bibfield  {author} {\bibinfo {author} {\bibfnamefont {R.~W.}\ \bibnamefont
  {Perry}}, \bibinfo {author} {\bibfnamefont {G.~N.}\ \bibnamefont {Meng}},
  \bibinfo {author} {\bibfnamefont {T.~G.}\ \bibnamefont {Dimiduk}}, \bibinfo
  {author} {\bibfnamefont {J.}~\bibnamefont {Fung}}, \ and\ \bibinfo {author}
  {\bibfnamefont {V.~N.}\ \bibnamefont {Manoharan}},\ }\href@noop {} {\bibfield
   {journal} {\bibinfo  {journal} {Faraday Discuss.}\ }\textbf {\bibinfo
  {volume} {159}},\ \bibinfo {pages} {211} (\bibinfo {year}
  {2012})}\BibitemShut {NoStop}%
\bibitem [{\citenamefont {Wang}\ \emph
  {et~al.}(2014{\natexlab{b}})\citenamefont {Wang}, \citenamefont {Dimiduk},
  \citenamefont {Fung}, \citenamefont {Razavi}, \citenamefont {Kretzschmar},
  \citenamefont {Chaudhary},\ and\ \citenamefont {Manoharan}}]{wang14using}%
  \BibitemOpen
  \bibfield  {author} {\bibinfo {author} {\bibfnamefont {A.}~\bibnamefont
  {Wang}}, \bibinfo {author} {\bibfnamefont {T.~G.}\ \bibnamefont {Dimiduk}},
  \bibinfo {author} {\bibfnamefont {J.}~\bibnamefont {Fung}}, \bibinfo {author}
  {\bibfnamefont {S.}~\bibnamefont {Razavi}}, \bibinfo {author} {\bibfnamefont
  {I.}~\bibnamefont {Kretzschmar}}, \bibinfo {author} {\bibfnamefont
  {K.}~\bibnamefont {Chaudhary}}, \ and\ \bibinfo {author} {\bibfnamefont
  {V.~N.}\ \bibnamefont {Manoharan}},\ }\href@noop {} {\bibfield  {journal}
  {\bibinfo  {journal} {J. Quant. Spectr. Rad. Trans.}\ }\textbf {\bibinfo
  {volume} {146}},\ \bibinfo {pages} {499} (\bibinfo {year}
  {2014}{\natexlab{b}})}\BibitemShut {NoStop}%
\bibitem [{\citenamefont {Yamada}, \citenamefont {Miyamoto},\ and\
  \citenamefont {Koizumi}(1985)}]{yamada85}%
  \BibitemOpen
  \bibfield  {author} {\bibinfo {author} {\bibfnamefont {Y.}~\bibnamefont
  {Yamada}}, \bibinfo {author} {\bibfnamefont {K.}~\bibnamefont {Miyamoto}}, \
  and\ \bibinfo {author} {\bibfnamefont {A.}~\bibnamefont {Koizumi}},\
  }\href@noop {} {\bibfield  {journal} {\bibinfo  {journal} {Aerosol Sci.
  Techn.}\ }\textbf {\bibinfo {volume} {4}},\ \bibinfo {pages} {227} (\bibinfo
  {year} {1985})}\BibitemShut {NoStop}%
\bibitem [{\citenamefont {Cermola}\ and\ \citenamefont
  {Schreil}(1987)}]{cermola87}%
  \BibitemOpen
  \bibfield  {author} {\bibinfo {author} {\bibfnamefont {M.}~\bibnamefont
  {Cermola}}\ and\ \bibinfo {author} {\bibfnamefont {W.-H.}\ \bibnamefont
  {Schreil}},\ }\href@noop {} {\bibfield  {journal} {\bibinfo  {journal}
  {Microscopy Res. Techn.}\ }\textbf {\bibinfo {volume} {5}},\ \bibinfo {pages}
  {171} (\bibinfo {year} {1987})}\BibitemShut {NoStop}%
\bibitem [{Note1()}]{Note1}%
  \BibitemOpen
  \bibinfo {note} {The ADDA implementation of the discrete-dipole approximation
  \cite {yurkin11} used for this study discretizes the particle volume on a
  three-dimensional square grid with an effective lattice constant roughly
  one-tenth the wavelength of light in the material. Typical numbers of dipoles
  range from \num {100} for the smallest particles considered to \num {18000}
  for the largest.}\BibitemShut {Stop}%
\end{thebibliography}
%

\end{document}